\documentclass[aps,pra,twocolumn,superscriptaddress,showpacs]{revtex4}
\usepackage{graphicx}
\usepackage{hyperref}

\begin{document}
\title{Experimental Generation of Large Quadrature EPR Entanglement with a Self-Phase-Locked Type II OPO Below Threshold}

\author{J. Laurat} \affiliation{Laboratoire
Kastler Brossel, UPMC, Case 74, 4 Place Jussieu, 75252 Paris cedex
05, France}
\author{T. Coudreau}\email{coudreau@spectro.jussieu.fr}\affiliation{Laboratoire
Kastler Brossel, UPMC, Case 74, 4 Place Jussieu, 75252 Paris cedex
05, France}\affiliation{Laboratoire Mat{\'e}riaux et Ph{\'e}nom{\`e}nes
Quantiques, Case 7021, Universit{\'e} Denis Diderot, 2 Place Jussieu,
75251 Paris cedex 05, France}
\author{G. Keller} \affiliation{Laboratoire
Kastler Brossel, UPMC, Case 74, 4 Place Jussieu, 75252 Paris cedex
05, France}
\author{N. Treps} \affiliation{Laboratoire
Kastler Brossel, UPMC, Case 74, 4 Place Jussieu, 75252 Paris cedex
05, France}
\author{C. Fabre } \affiliation{Laboratoire
Kastler Brossel, UPMC, Case 74, 4 Place Jussieu, 75252 Paris cedex
05, France}

\date{\today}

\begin{abstract}
We study theoretically and experimentally the quantum properties
of a type II frequency degenerate optical parametric oscillator
below threshold with a quarter-wave plate inserted inside the
cavity which induces a linear coupling between the orthogonally
polarized signal and idler fields. This original device provides a
good insight into general properties of two-mode gaussian states,
illustrated in terms of covariance matrix. We report on the
experimental generation of two-mode squeezed vacuum on
non-orthogonal quadratures depending on the plate angle. After a
simple operation, the entanglement is maximized and put into
standard form, \textit{i.e.} quantum correlations and
anti-correlations on orthogonal quadratures. A half-sum of
squeezed variances as low as $0.33 \pm 0.02$, well below the unit
limit for inseparability, is obtained and the entanglement
measured by the entropy of formation.
\end{abstract}

 \maketitle

\section{Introduction}
\label{sec:intro} The dynamic and promising field of quantum
information with continuous variables aroused a lot of interest
and a large number of protocols has been proposed and implemented
\cite{CV}. Continuous variable entanglement plays a central role
and constitutes the basic requisite of most of these developments.
Such a ressource can be generated by mixing on a beam-splitter two
independent squeezed beams produced for instance by type-I OPAs
\cite{Furusawa,Bowen} or by Kerr effect in a fiber
\cite{Silberhorn}. The use of a light field interacting with a
cloud of cold atoms in cavity has also been recently reported
\cite{josse}. Another way is to use a type-II OPO below threshold
-- with vacuum \cite{Ou,Schori} or coherent injection \cite{Zhang}
-- which directly provides orthogonally polarized entangled beams.

We propose here to explore the quantum properties of an original
device -- called a "self-phase-locked OPO" -- which consists of a
type-II OPO with a quarter-wave plate inserted inside the cavity
\cite{mason}. The plate -- which can be rotated relative to the
principal axis of the crystal -- adds a linear coupling between
the orthogonally polarized signal and idler fields. It has been
shown that such a device above threshold opens the possibility to
produce frequency degenerate bright EPR beams thanks to the
phase-locking resulting from the linear coupling induced by the
rotated plate \cite{longcham1,longcham2}. Such a device can also
be operated below threshold and exhibits a very rich quantum
behavior. The paper is devoted to this below threshold regime from
a theoretical and experimental point of view. The properties are
interpreted in terms of covariance matrix and give an interesting
insight into the non-classical properties of two-mode gaussian
states -- such as squeezing, entanglement and their respective
links. The strongest EPR entanglement to date is then reported.

The paper is organized as follows. In Sec.\ref{sec:theory}, we
describe the quantum state generated by a self-phase-locked type
II OPO below threshold. The correlated quadratures and the amount
of entanglement depend on the angle of the wave-plate. Different
regimes are identified and a necessary operation to maximize
entanglement is described and interpreted in terms of covariance
matrix and logarithmic negativity. The experimental setup is
presented in Sec.\ref{sec:setup} and a detection scheme relying on
two simultaneous homodyne detections is detailed. Section
\ref{sec:entanglement} is devoted to the experimental results. In
Sec.\ref{sec:conclusion}, the main conclusions of the experimental
work are summarized and the extension to the above threshold
regime discussed.

\section{Theory of self-phase-locked OPO below threshold}
\label{sec:theory} In this section, we present a theoretical
analysis of the quantum properties of the self-phase-locked OPO
below threshold by the usual linearization technique
\cite{reynaud}. Individual quantum noise properties of the signal
and idler modes as well as their correlations are derived.

\subsection{Linearized equations with linear coupling}
\begin{figure}
\includegraphics[width=.9\columnwidth,clip=]{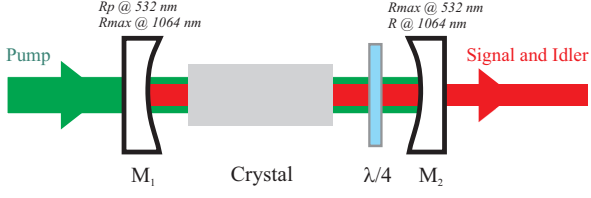}%
\caption{Linear cavity OPO with a quarter-wave plate. When rotated
relatively to the principal axes of the type-II phase-matched
crystal, this plate adds a linear coupling between orthogonally
polarized signal and idler fields. Vacuum fluctuations can enter
the system through the output mirror or the various
losses.}\label{opo}
\end{figure}

The self-phase-locked type II OPO is sketched in Fig. \ref{opo}. A
quarter-wave plate and a type-II phase matched $\chi^{(2)}$
crystal are both inserted inside a triply resonant linear cavity.
The plate can be rotated by an angle $\rho$ with respect to the
crystal neutral axes. In this paper, we will restrict ourselves to
small values of $\rho$.

The damping rate is assumed to be the same for the signal and
idler modes. As the finesse is high, we note $r=1-\kappa$ the
amplitude reflection coefficient for these modes, with $\kappa \ll
1$. The intensity transmission coefficient is thus approximatively
equal to $2\kappa$. To take into account the additional losses
undergone by the signal and idler modes (crystal absorption,
surface scattering), we introduce a generalized reflection
coefficient $r'=1-\kappa'=1-(\kappa+\mu$). For the sake of
simplicity, all coefficients are assumed to be real and the phase
matching will be taken perfect. The influence of different
reflection phase-shifts on the cavity mirrors for the interacting
waves has been detailed in Ref. \cite{longcham1} in the above
threshold regime and will not be considered here.

We assume that the signal and idler modes are close to resonance
and note $\Delta_{1}$ and $\Delta_{2}$ their small round trip
phase detunings. The equations of motion for the classical field
amplitudes -- which are noted $A_{1}$ and $A_{2}$ for the signal
and idler modes and $A_{0}$ for the pump -- can be written as
\begin{eqnarray}
\tau \frac{dA_{1}}{dt}&=&A_{1}(-\kappa'+i\Delta_{1})+gA_{0}A_{2}^{\ast}+2i\rho e^{i(\theta-\psi)}A_{2} \nonumber \\
\tau
\frac{dA_{2}}{dt}&=&A_{2}(-\kappa'+i\Delta_{2})+gA_{0}A_{1}^{\ast}+2i\rho
e^{i(\psi-\theta)}A_{1}
\end{eqnarray}
where $\tau$ stands for the cavity round-trip time, $A_{0}^{in}$
for the input pump amplitude and $g$ for the parametric gain.
$\theta$ and $\psi$ are respectively the birefringent phase shift
introduced by the crystal and by the waveplate. The last term of
these equations corresponds to the linear coupling induced by the
rotated plate.

We will only consider the case where $\Delta_{1}=\Delta_{2}=2\rho$
and $\theta=\psi$. At this operating point the threshold is
minimum \cite{longcham1}. In this case, the equations of motion
are simpler and are written
\begin{eqnarray}
\tau \frac{dA_{1}}{dt}&=&A_{1}(-\kappa'+2i\rho)+gA_{0}A_{2}^{\ast}+2i\rho A_{2} \nonumber \\
\tau
\frac{dA_{2}}{dt}&=&A_{2}(-\kappa'+2i\rho)+gA_{0}A_{1}^{\ast}+2i\rho
A_{1}
\end{eqnarray}

A non-zero stationary solution exists if and only if the pump
power $A_{0}$, taken real, exceeds the threshold power equal to
$\frac{\kappa'}{g}$. We define a reduced pumping parameter
$\sigma$ equal to the input pump amplitude normalized to the
threshold. The below threshold regime corresponds to $\sigma<1$.

These equations are linearized around the stationary values by
setting $A_{i}=\overline{A_{i}}+\delta A_{i}$. In the below
threshold regime, the mean value of $A_{1}$ and $A_{2}$ are zero.
The linearized equations can then be written

\begin{eqnarray}
-\frac{\tau}{\kappa'} \frac{d(\delta A_{1})}{dt}&=&\delta A_{1}(1-ic)-\sigma \delta A_{2}^{\ast}-ic \delta A_{2}\nonumber \\&& -\frac{\sqrt{2\kappa}}{\kappa'}\delta A_{1}^{in}-\frac{\sqrt{2\mu}}{\kappa'}\delta B_{1}^{in} \nonumber \\
-\frac{\tau}{\kappa'} \frac{d(\delta A_{2})}{dt}&=&\delta
A_{2}(1-ic)-\sigma \delta A_{1}^{\ast}-ic \delta A_{1}\nonumber
\\&& -\frac{\sqrt{2\kappa}}{\kappa'}\delta
A_{2}^{in}-\frac{\sqrt{2\mu}}{\kappa'}\delta B_{2}^{in}
\end{eqnarray}
where $c=\frac{2\rho}{\kappa'}$. $\delta A_{i}^{in}$ and $\delta
B_{i}^{in}$ correspond to the vacuum fluctuations entering the
cavity due respectively to the coupling mirror and to the losses.

One can note that the fluctuations of the pump are not coupled to
the signal and idler modes in the below threshold regime. It is
obviously not the case above threshold and this point can explain
in particular why the experimental observation above threshold of
phase anti-correlations below the standard quantum limit is a
difficult task \cite{laurat04d}.

\subsection{Variances}
The fluctuations can be evaluated by taking the Fourier transform
of the previous equations which leads to algebraic equations. We
introduce the parameter
$\Omega=\frac{\omega\tau}{2\kappa'}=\frac{\omega}{\Omega_{c}}$,
which is the noise frequency normalized to the cavity bandwidth
$\Omega_{c}$. In the Fourier domain, the equations become
\begin{eqnarray}
(1-ic+2i\Omega)\delta \widetilde{A_{1}}(\Omega)-\sigma \delta \widetilde{A_{2}^{\ast}}(-\Omega)-ic \delta \widetilde{A_{2}}(\Omega)-&&\nonumber \\\frac{\sqrt{2\kappa}}{\kappa'} \delta \widetilde{A_{1}^{in}}(\Omega)-\frac{\sqrt{2\mu}}{\kappa'}\delta \widetilde{B_{1}^{in}}(\Omega)=0\qquad \nonumber \\
(1-ic+2i\Omega)\delta \widetilde{A_{2}}(\Omega)-\sigma \delta
\widetilde{A_{1}^{\ast}}(-\Omega)-ic \delta
\widetilde{A_{1}}(\Omega)-&&\nonumber
\\\frac{\sqrt{2\kappa}}{\kappa'} \delta
\widetilde{A_{2}^{in}}(\Omega)-\frac{\sqrt{2\mu}}{\kappa'}\delta
\widetilde{B_{2}^{in}}(\Omega)=0\qquad
\end{eqnarray}

From these equations and their conjugates, one can determine the
variance spectra of the signal and idler modes and their
correlations. We define the fluctuations of the modes $A_{i}$ for
a given quadrature angle $\varphi$ and a given noise frequency
$\Omega$ by
\begin{eqnarray}
p_{i}(\varphi)=\delta
\widetilde{A_{i}}(\Omega)e^{-i\varphi}+\delta
\widetilde{A_{i}^{\ast}}(-\Omega)e^{i\varphi}
\end{eqnarray}

The equation of motion for the fluctuations can thus take the
following form
\begin{eqnarray}
(1+2i\Omega) p_{1}(\varphi)-\sigma p_{2}(-\varphi)+c (p_{1}(\varphi+\pi/2)\nonumber\\+p_{2}(\varphi+\pi/2))-\frac{\sqrt{2\kappa}}{\kappa'} p_{1}^{in}-\frac{\sqrt{2\mu}}{\kappa'}p_{1}^{in'}&=&0 \nonumber \\
p_{2}(-\varphi)(1+2i\Omega)-\sigma p_{1}(\varphi)+c
(p_{2}(-\varphi+\pi/2)\nonumber\\-p_{1}(-\varphi+\pi/2))-\frac{\sqrt{2\kappa}}{\kappa'}
p_{2}^{in}-\frac{\sqrt{2\mu}}{\kappa'}p_{2}^{in'}&=&0
\end{eqnarray}
where $p_{i}^{in}$ and $p_{i}^{in'}$ correspond to the
phase-insensitive vacuum fluctuations entering the system.

When $c=0$, these equations are identical to the ones of a
traditional type II OPO below threshold where only the quadratures
with phases $\pm\varphi$ can interact. When the plate is rotated,
the phase dependence becomes more complicated since orthogonal
quadratures are coupled.

\begin{figure}[htpb!]
\includegraphics[width=.95\columnwidth]{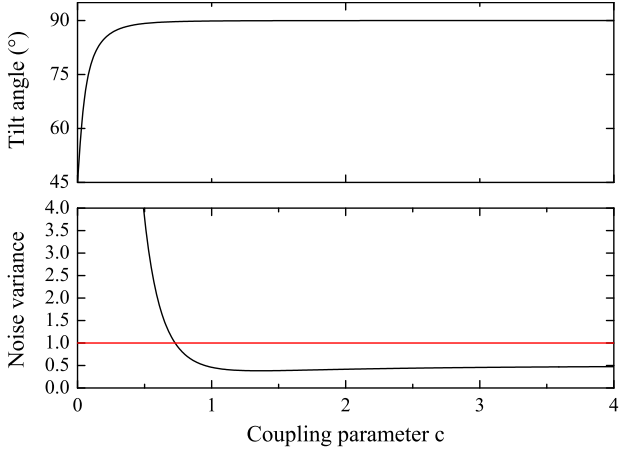}%
\caption{\label{individual} Angle of the minimal noise quadrature
and corresponding normalized variance as a function of the
coupling parameter. Close to $c=0$, the noise diverges and is
phase-insensitive. $\Omega=0, \sigma=0.9, \kappa=\kappa'$.}
\end{figure}

By introducing the simplified notations
\begin{eqnarray}
&p_{1}=p_{1}\Big(\varphi_{1}\Big)& \qquad
q_{1}=p_{1}\Big(\varphi_{1}+\frac{\pi}{2}\Big) \nonumber \\
&p_{2}=p_{2}\Big(\varphi_{2}\Big)& \qquad
q_{2}=p_{2}\Big(\varphi_{2}+\frac{\pi}{2}\Big)
\end{eqnarray}
with $\varphi_1=\pi/2$ and $\varphi_2=-\pi/2$, the equations of
motion can be rewritten in the form
\begin{eqnarray}
(1+2i\Omega)p_{1}-\sigma p_{2}+c
(q_{1}-q_{2})\nonumber\\-\frac{\sqrt{2\kappa}}{\kappa'}
p_{1}^{in}-\frac{\sqrt{2\mu}}{\kappa'}p_{1}^{in'}&=&0\nonumber\\
p_{2}(1+2i\Omega)-\sigma p_{1}+c
(q_{2}-q_{1})\nonumber\\-\frac{\sqrt{2\kappa}}{\kappa'}
p_{2}^{in}-\frac{\sqrt{2\mu}}{\kappa'}p_{2}^{in'}&=&0\nonumber\\
q_{1}(1+2i\Omega)+\sigma q_{2}+c
(p_{2}-p_{1})\nonumber\\-\frac{\sqrt{2\kappa}}{\kappa'}
q_{1}^{in}-\frac{\sqrt{2\mu}}{\kappa'}q_{1}^{in'}&=&0\nonumber\\
q_{2}(1+2i\Omega)+\sigma q_{1}+c
(p_{1}-p_{2})\nonumber\\-\frac{\sqrt{2\kappa}}{\kappa'}
q_{2}^{in}-\frac{\sqrt{2\mu}}{\kappa'}q_{2}^{in'}&=&0
\end{eqnarray}
The system made up of these 4 equations and the 4 equations
obtained by changing $\Omega$ in $-\Omega$ gives the intra-cavity
fluctuations. The fluctuations of the output modes are obtained by
the boundary condition on the output mirror
\begin{eqnarray}
p_{i}^{out}(\Omega)=\sqrt{2\kappa}p_{i}(\Omega)-p_{i}^{in}
\end{eqnarray}
The variances of a component $p_{i}^{out}$ is then derived from
\begin{eqnarray}
S_{p_{i}^{out}}(\Omega)=<p_{i}^{out}(\Omega)\,p_{i}^{out}(-\Omega)>
\end{eqnarray}
The variances of the uncorrelated vacuum contributions entering
the system are normalized to 1.

\subsection{Signal and idler
fluctuations} When the plate is not rotated ($c=0$), the signal
and idler modes exhibit phase-insensitive excess noise. The
single-beam noise spectrum for the signal (or the idler) can be
written
\begin{eqnarray}
S_{p_{1}}=S_{q_{1}}=1+\frac{8\sigma^{2}}{(4\Omega^{2}+(\sigma-1)^{2})(4\Omega^{2}+(\sigma+1)^{2})}\frac{\kappa}{\kappa'}
\end{eqnarray}

It is not the case when the plate is rotated. The noise becomes
phase-sensitive and the noise spectrum is given by
\begin{eqnarray}
S(\varphi)=S_{p_{1}}\cos(\varphi)^{2}+S_{q_{1}}\sin(\varphi)^{2}+\alpha
\cos(2\varphi)
\end{eqnarray}
with
\begin{widetext}
\begin{eqnarray}
S_{p_{1}}&=&1+\frac{8\sigma(\sigma((\sigma-1)^{2}+4\Omega^{2})-c^{2}(4\Omega^{2}-4(1+c^{2})+(\sigma+1)^{2})}{(4\Omega^{2}+(\sigma-1)^{2})(16\Omega^{2}+(4\Omega^{2}-4c^{2}+\sigma^{2}-1)^{2})}\frac{\kappa}{\kappa'}\nonumber\\
S_{q_{1}}&=&1+\frac{8\sigma(\sigma((\sigma+1)^{2}+4\Omega^{2})+c^{2}(4\Omega^{2}-4(1+c^{2})+(\sigma-1)^{2})}{(4\Omega^{2}+(\sigma+1)^{2})(16\Omega^{2}+(4\Omega^{2}-4c^{2}+\sigma^{2}-1)^{2})}\frac{\kappa}{\kappa'}\nonumber\\
\alpha&=&\frac{-8\sigma
c}{16\Omega^{2}+(4\Omega^{2}-4c^{2}+\sigma^{2}-1)^{2}}\frac{\kappa}{\kappa'}
\label{alpha}\end{eqnarray}
\end{widetext}

Figure \ref{individual} shows the evolution of the minimal noise
quadrature angle and of the corresponding noise power as a
function of the coupling parameter. When the coupling parameter
$c$ increases, this quadrature rotates. For strong coupling, the
minimal noise quadrature is closer and closer to the quadrature
$q_{1}$ and $q_{2}$ and the noise can be squeezed well below the
standard quantum limit.

\subsection{Correlations and anti-correlations}
After considering the individual fluctuations of signal and idler
modes, we study here the intermodal correlations.

Let us introduce the superposition modes oriented $\pm45^{\circ}$
from the axes of the crystal
\begin{eqnarray} A_{+}=\frac{A_{1}+A_{2}}{\sqrt{2}} \qquad
\textrm{and} \qquad A_{-}=\frac{A_{1}-A_{2}}{\sqrt{2}} \nonumber
\end{eqnarray}
It should be stressed that considering the noise spectrum of the
sum or difference of signal and idler fluctuations is equivalent
to considering the noise spectrum of the rotated modes. If signal
and idler exhibit correlations or anti-correlations, these two
modes can have squeezed fluctuations as their noise spectra are
given by
\begin{eqnarray} S_{A_{+}}(\varphi)=\frac{1}{2}S_{p_{1}(\varphi)+p_{2}(\varphi)} \quad
\textrm{and} \quad
S_{A_{-}}(\varphi)=\frac{1}{2}S_{p_{1}(\varphi)-p_{2}(\varphi)}
\nonumber
\end{eqnarray}
The amount of entanglement between signal and idler can be
directly inferred from the amount of squeezing available on these
superposition modes.

The expressions for the anti-correlations between signal and idler
modes coincide with the ones obtained in the case of a standard
OPO below threshold
\begin{eqnarray}\label{without}
S_{q_{1}+q_{2}}&=&1-\frac{4\sigma}{4\Omega^{2}+(\sigma+1)^{2}}\frac{\kappa}{\kappa'}\nonumber\\
S_{p_{1}+p_{2}}&=&1+\frac{4\sigma}{4\Omega^{2}+(\sigma-1)^{2}}\frac{\kappa}{\kappa'}
\end{eqnarray}
The combination $(q_{1}+q_{2})$ is always squeezed below the
standard quantum limit while $(p_{1}+p_{2})$ is very noisy.
Perfect anti-correlations are found at exact threshold in the
absence of additional losses ($\kappa=\kappa'$) and at zero
frequency.

In contrast with the anti-correlations, the correlations largely
depend on the presence of the plate. The variance spectrum is
found to be
\begin{eqnarray}\label{amoins1}
S_{A_{-}}(\varphi)=S_{p_{1}-p_{2}}\cos(\varphi)^{2}+S_{q_{1}-q_{2}}\sin(\varphi)^{2}+2\alpha\cos(2\varphi)
\end{eqnarray}
where $\alpha$ has been defined in Eq. (\ref{alpha}) and
\begin{eqnarray}\label{amoins2}
S_{p_{1}-p_{2}}&=&1-\frac{4\sigma(4\Omega^{2}-4c^{2}+(\sigma-1)^{2})}{16\Omega^{2}+(4\Omega^{2}-4c^{2}+\sigma^{2}-1)}\frac{\kappa}{\kappa'}\nonumber\\
S_{q_{1}-q_{2}}&=&1+\frac{4\sigma(4\Omega^{2}-4c^{2}+(\sigma+1)^{2})}{16\Omega^{2}+(4\Omega^{2}-4c^{2}+\sigma^{2}-1)}\frac{\kappa}{\kappa'}
\end{eqnarray}

\begin{figure}[htpb!]
\includegraphics[width=.95\columnwidth]{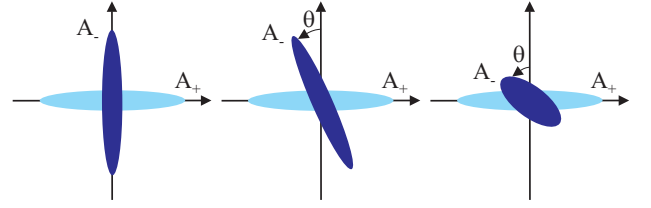}%
\caption{\label{ellipse} Fresnel representation of the noise
ellipse of the $\pm 45^{\circ}$ rotated modes when the plate angle
is increased. Without coupling, squeezing is predicted on
orthogonal quadratures. The noise ellipse of the $-45^{\circ}$
mode rotates and the noise reduction is degraded when the coupling
increases while the $+45^{\circ}$ rotated mode is not affected.}
\end{figure}

In a standard OPO below threshold -- without a linear coupling --
the correlated quadratures are orthogonal to the anti-correlated
ones. It is not anymore the case when a coupling is introduced.
The evolution is depicted in Fig. \ref{ellipse}. When the plate
angle increases, the correlated quadratures rotates and the
correlations are degraded.

We can derive from Eqs (\ref{amoins1}) and (\ref{amoins2}) a
simple expression for the tilt angle $\theta$ of the noise ellipse
\begin{eqnarray}\label{theta}
\tan(2\theta)=\frac{4c}{4\Omega^{2}-4c^{2}+\sigma^{2}+1}
\end{eqnarray}

\begin{figure}[htpb!]
\includegraphics[width=.95\columnwidth]{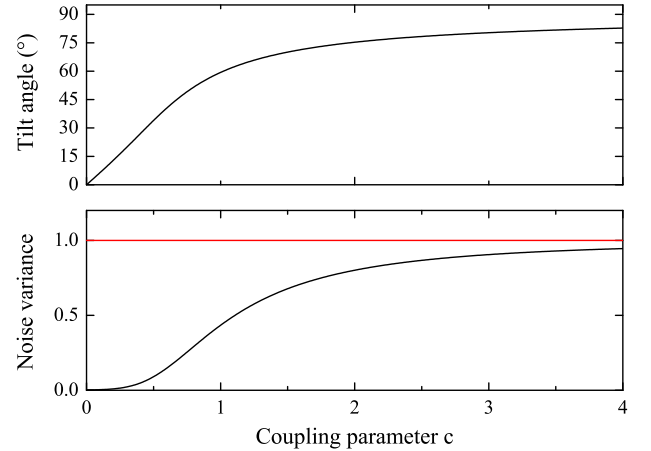}%
\caption{\label{Amoins} Angle $\theta$ of the minimal noise
quadrature and corresponding normalized variance as a function of
the coupling parameter $c$. $\Omega=0, \sigma=0.9,
\kappa=\kappa'$. }
\end{figure}

Figure \ref{Amoins} gives the tilt angle of the noise ellipse and
the noise variance of the squeezed quadrature as a function of the
coupling parameter.

As a first conclusion, optimal correlations and anti-correlations
are observed on non-orthogonal quadratures depending on the plate
angle. In order to maximize the entanglement between the signal
and idler modes, the optimal quadratures have to be made
orthogonal \cite{Wolf}. Such an operation consists in a
phase-shift of $A_{-}$ relative to $A_{+}$. This transformation is
thus "non-local" in the sense of the EPR argument: it involves the
two considered modes, signal and idler, and therefore has to be
performed before spatially separating them.

\begin{figure*}[htpb!]
\small
\begin{eqnarray}
\Gamma_{A_{1}\,\!A_{2}} = \left( \begin{array}{cc|cc}
181.192 & 0 & 179.808 & -0.255 \\
0 & 0.386 & -0.255 & -0.383 \\
\hline 179.808 & -0.255 & 181.192 & 0\\
-0.255 & -0.383 & 0 & 0.386
\end{array} \right)
&\Longrightarrow& \Gamma'_{A_{1}\,\!A_{2}} = \left(
\begin{array}{cc|cc}
180.839 & 0 & 180.161 & 0 \\
0 & 0.739 & 0 & -0.736 \\
\hline 180.161 & 0 & 180.839 & 0\\
0 & -0.736 & 0 & 0.739
\end{array} \right)
\nonumber
\end{eqnarray}
\begin{eqnarray}
\Gamma_{A_{+}\,\!A_{-}} = \left( \begin{array}{cc|cc}
361 & 0 & 0 & 0 \\
0 & 0.00277 & 0 & 0 \\
\hline 0& 0 & 1.383 & -0.256\\
0 & 0 & -0.256 & 0.770
\end{array} \right)
&\Longrightarrow& \Gamma'_{A_{+}\,\!A_{-}} = \left(
\begin{array}{cc|cc}
361 & 0 & 0 & 0 \\
0 & 0.00277 & 0 & 0 \\
\hline 0& 0 & 0.677 & 0\\
0 & 0 & 0 & 1.476
\end{array} \right)\nonumber
\end{eqnarray}
\caption{Numerical example of covariance matrix of the
$A_{1}$/$A_{2}$ modes and the $A_{+}$/$A_{-}$ modes before and
after the non-local operation for a coupling parameter $c=1.5$.
($\sigma=0.9, \Omega=0, \kappa=\kappa'$)} \label{matrix}
\end{figure*}
\subsection{In terms of covariance matrix}

The behavior of the system and the optimization of the degree of
entanglement can be formulated in terms of covariance matrix. We
recall that a two-mode gaussian state with zero mean value is
fully described by the covariance matrix $\Gamma_{A\,\!B}$ defined
as
\begin{equation}
\Gamma_{A\,\!B}=\left( \begin{array}{cc}
 \gamma_{A} & \sigma_{A\,\!B} \\
 \sigma_{A\,\!B}^{T} & \gamma_{B}
\end{array} \right)\nonumber
\end{equation}
$\gamma_{A}$ and $\gamma_{B}$ are the covariance matrix of the
individual modes while $\sigma_{A\,\!B}$ describes the intermodal
correlations. The elements of the covariance matrix are written
$\Gamma_{ij}=\langle \delta R_{i}\delta R_{j}+\delta R_{j}\delta
R_{i}\rangle/2$ where
$R_{\{i,i=1,..,4\}}=\{X_{A},Y_{A},X_{B},Y_{B}\}$. $X$ and $Y$
corresponds to an arbitrary orthogonal basis of quadratures.

In order to measure the degree of entanglement of Gaussian states,
a simple computable formula of the logarithmic negativity
$E_{\mathcal{N}}$ has been obtained in Ref. \cite{vidal} (see also
\cite{Adesso} for a general overview). $E_{\mathcal{N}}$ can be
easily evaluated from the largest positive symplectic eigenvalue
$\xi$ of the covariance matrix which can be obtained from
\begin{eqnarray}
\xi^{2}=\frac{1}{2}(D-\sqrt{D^{2}-4\det\Gamma_{A\,\!B}}\,)
\end{eqnarray}
with
\begin{eqnarray}
D=\det\gamma_{A}+\det\gamma_{B}-2\det\sigma_{A\,\!B}
\end{eqnarray}
The two-mode state is entangled if and only if $\xi<1$. The
logarithmic negativity can thus be expressed by
$E_{\mathcal{N}}=-\log_{2}(\xi)$. This measurement is monotone and
can not increase under LOCC (local operations and classical
communications). The maximal entanglement which can be extracted
from a given two-mode state by passive operations is related to
the two smallest eigenvalues of $\Gamma$, $\lambda_1$ and
$\lambda_2$, by
$E_{\mathcal{N}}^{max}=-\log_{2}(\lambda_1\lambda_2)/2$
\cite{Wolf}.

Phase-shifting of $A_{+}$ and $A_{-}$ into $A_{+}\,e^{i\theta/2}$
and $A_{-}\,e^{-i\theta/2}$ corresponds to a transformation of the
signal and idler modes $A_{1}$ and $A_{2}$ described by the matrix
\begin{displaymath}
M = \left( \begin{array}{cc}
\cos(\theta/2) & i \sin(\theta/2) \\
i \sin(\theta/2) & \cos(\theta/2) \\
\end{array} \right)
\end{displaymath}
The angle $\theta$ is given by Eq.(\ref{theta}). Such a
transformation couples the signal and idler modes.

We give here a numerical example for realistic experimental values
$c=1.5, \sigma=0.9$ and $\Omega=0$. The covariance matrix for the
$A_{1}$/$A_{2}$ modes and also for the $A_{+}$/$A_{-}$ modes are
given in Fig. \ref{matrix} with and without the phase-shift. The
matrix of the $A_{+}$/$A_{-}$ modes are well-suited to understand
the behavior of the device. At first, the intermodal blocks are
zero, showing that these two modes are not at all correlated and
consequently are the most squeezed modes of the system. There is
no way to extract more squeezing. But one can also note that the
diagonal blocks are not diagonalized simultaneously. This
corresponds to the tilt angle $\theta$ of the squeezed quadrature
of $A_{-}$ and given by Eq. (\ref{theta}). A phase-shift of the
angle $\theta$ permits to diagonalize simultaneously the two
blocks and to obtain squeezing on orthogonal quadratures. From the
matrix on the $A_{1}$/$A_{2}$ modes, one can quantify the degree
of entanglement by the logarithmic negativity $E_{\mathcal{N}}$.
Thanks to the non-local operation, $E_{\mathcal{N}}$ goes from
$4.06$ to $4.53$. The maximal entanglement available has been
extracted as
$E_{\mathcal{N}}^{max}=-\log_{2}(\lambda_1\lambda_2)/2=4.53$. Let
us finally note that due to the strong coupling the signal and
idler
modes are entangled but also slightly squeezed.\\

A self-phase locked OPO below threshold can generate very strong
entangled modes when the plate angle is small enough. The quantum
behavior of the device is very rich and gives a good insight into
two-mode gaussian state properties and entanglement
characterization. The previous interpretation in terms of
covariance matrix establishes a link between the optimal
entanglement that can be extracted and the eigenvalues of the
matrix. The way to find it by a non-local operation is developed.
The next section is devoted to the experimental study of this
original device.

\begin{figure*}
\includegraphics[width=1.9\columnwidth]{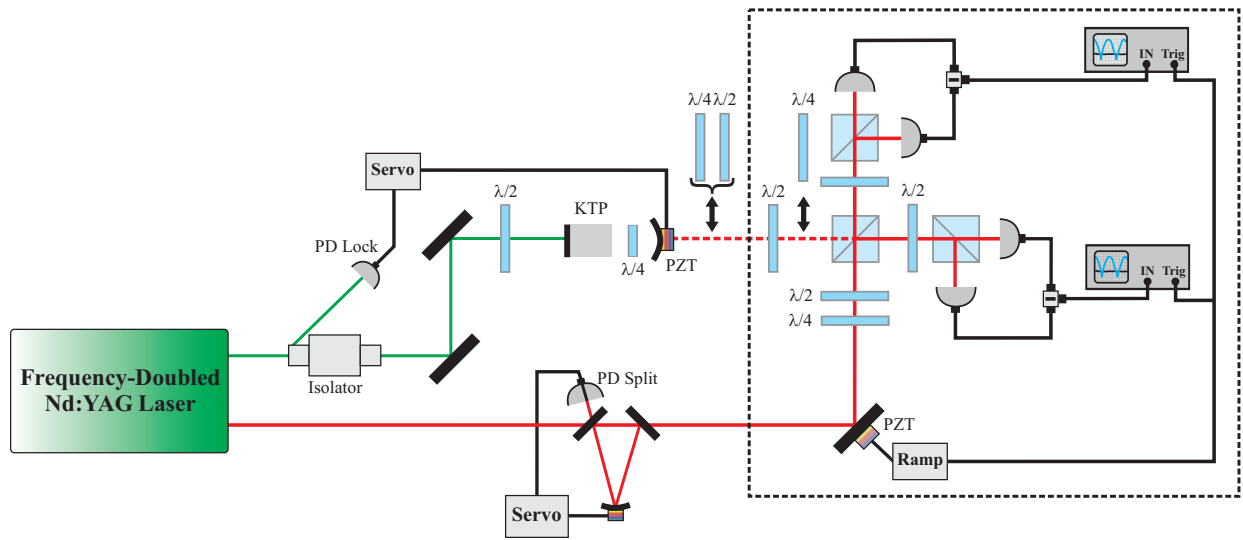}
\caption{Experimental setup. A continuous-wave frequency-doubled
Nd:YAG laser pumps below threshold a type II OPO with a
$\lambda/4$ plate inserted inside the cavity. The generated
two-mode vacuum state is characterized by two simultaneous
homodyne detections. The infrared output of the laser is used as
local oscillator after filtering by a high-finesse cavity. The two
couples $\{\lambda/4,\lambda/2\}$ on each path are used to make
arbitrary phase shift between orthogonal components of
polarization. PD Lock: FND-100 photodiode for locking of the OPO.
PD Split: split two-element InGaAs photodiode for tilt-locking of
the filtering cavity.}\label{setup}
\end{figure*}

\section{Experimental setup}
\label{sec:setup} Our experimental setup is based on a frequency
degenerate type II OPO below threshold. A $\lambda/4$ plate
inserted within the OPO adds a linear coupling between the signal
and idler modes which depends on the angle of the plate relative
to the principal axes of the crystal . Two simultaneous homodyne
detections are implemented.

\subsection{OPO and linear coupling}

The experimental setup is shown in Fig. \ref{setup}. A continuous
frequency-doubled Nd:YAG laser ("Diabolo" without "noise eater
option", Innolight GmbH) pumps a triply resonant  type II OPO,
made of a semi-monolithic linear cavity : in order to increase the
mechanical stability and reduce the reflection losses, the input
flat mirror is directly coated on one face of the 10mm-long KTP
crystal ($\theta=90^{\circ}$, $\varphi=23.5^{\circ}$, Raicol
Crystals Ltd.). The reflectivities for the input coupler are 95\%
for the pump (532nm) and almost 100\% for the signal and idler
beams (1064nm). The output coupler (R=38mm) is highly reflecting
for the pump and its transmission is 5\% for the infrared. At
exact triple resonance, the oscillation threshold is less than 20
mW, very close to the threshold without the plate \cite{laurat03}.
The OPO is actively locked on the pump resonance by the
Pound-Drever-Hall technique: a remaining 12MHz modulation present
in the laser is detected by reflection and the error signal is
sent to a home-made PI controller. The crystal temperature is
thoroughly controlled within the mK range. The OPO can operate
stably during more than one hour without mode-hopping.

The birefringent plate inserted inside the cavity is chosen to be
exactly $\lambda/4$ at 1064 nm and almost $\lambda$ at the 532 nm
pump wavelength. As birefringence and dispersion are of the same
order, this configuration is only possible by choosing
multiple-order plate: we have chosen the first order for which
exact $\lambda/4$ at 1064 nm is obtained, i.e. $4.75 \lambda$ at
1064 nm and $9.996\lambda$ at 532 nm. Very small rotations of this
plate around the cavity axis can be operated thanks to a rotation
mount controlled by piezo-electric actuator (New Focus Model 8401
and tiny pico-motor).

\subsection{Two simultaneous homodyne detections}

The coherent 1064 nm laser output is used as local oscillator for
homodyne detection. This beam is spatially filtered and
intensity-noise cleaned by a triangular-ring 45 cm-long cavity
with a high finesse of 3000. This cavity is locked on the maximum
of transmission by the single-pass tilt-locking technique
\cite{Shaddock} and 80\% of transmission is obtained. The homodyne
detections are based on pairs of balanced high quantum efficiency
InGaAs photodiodes (Epitaxx ETX300 with a 95\% quantum efficiency)
and the fringe visibility reaches 0.97. The shot noise level of
all measurements is easily obtained by blocking the output of the
OPO.

Orthogonally polarized modes are separated on the first polarizing
beam splitter at the output of the OPO. A half-wave plate inserted
before this polarizing beam splitter enables us to choose the
fields to characterize: the signal and idler modes which are
entangled, or the $\pm45^{\circ}$ rotated modes which are
squeezed.

One main requirement of our experiment is to be able to
characterize simultaneously two modes with the same phase
reference. The difference photocurrents of the homodyne detections
are sent into two spectrum analyzers (Agilent E4411B) which are
triggered by the same signal. The two homodyne detections are
calibrated in order to be in phase: if one send into each
detection a state of light with squeezing on the same quadrature,
the noise powers registered on the spectrum analyzers must have
in-phase variations while scanning the local oscillator phase. Two
birefringent plates, $\lambda/2$ and $\lambda/4$, inserted in the
local oscillator path are rotated in order to compensate residual
birefringence due in particular to defects associated to
polarizing beam splitter. In others words, after this correction,
the polarization of the local oscillator is slightly elliptical.
To facilitate this tuning, the OPO is operated above threshold in
the locking zone where frequency degeneracy occurs. A polarizing
beam splitter inserted at the OPO output and a $\lambda/2$ plate
rotated by $22.5^{\circ}$ permit to send into the two homodyne
detections states of light with opposite phase. Then, we look at
the DC interference fringes which have to be in opposition. We
check this calibration by sending into the homodyne detections a
squeezed vacuum. When scanning the local oscillator phase, the
noise variance measured in each homodyne detection follow
simultaneous variations. A $\lambda/4$ plate can be added on the
beam exiting the OPO, just before the homodyne detections: when
this plate is inserted, the homodyne detections are in quadrature.
In such a configuration, two states of light with squeezing on
orthogonal quadratures give in-phase squeezing curves on the
spectrum analyzers.

\section{Experimental entanglement}
\label{sec:entanglement} In this section, we report on the
experimental results obtained for different values of the coupling
parameter. As underlined before, we characterize the noise of the
$\pm 45^{\circ}$ rotated modes which have squeezed fluctuations.

\subsection{Without linear coupling}
\begin{figure}[htpb!]
\includegraphics[width=.95\columnwidth,clip=]{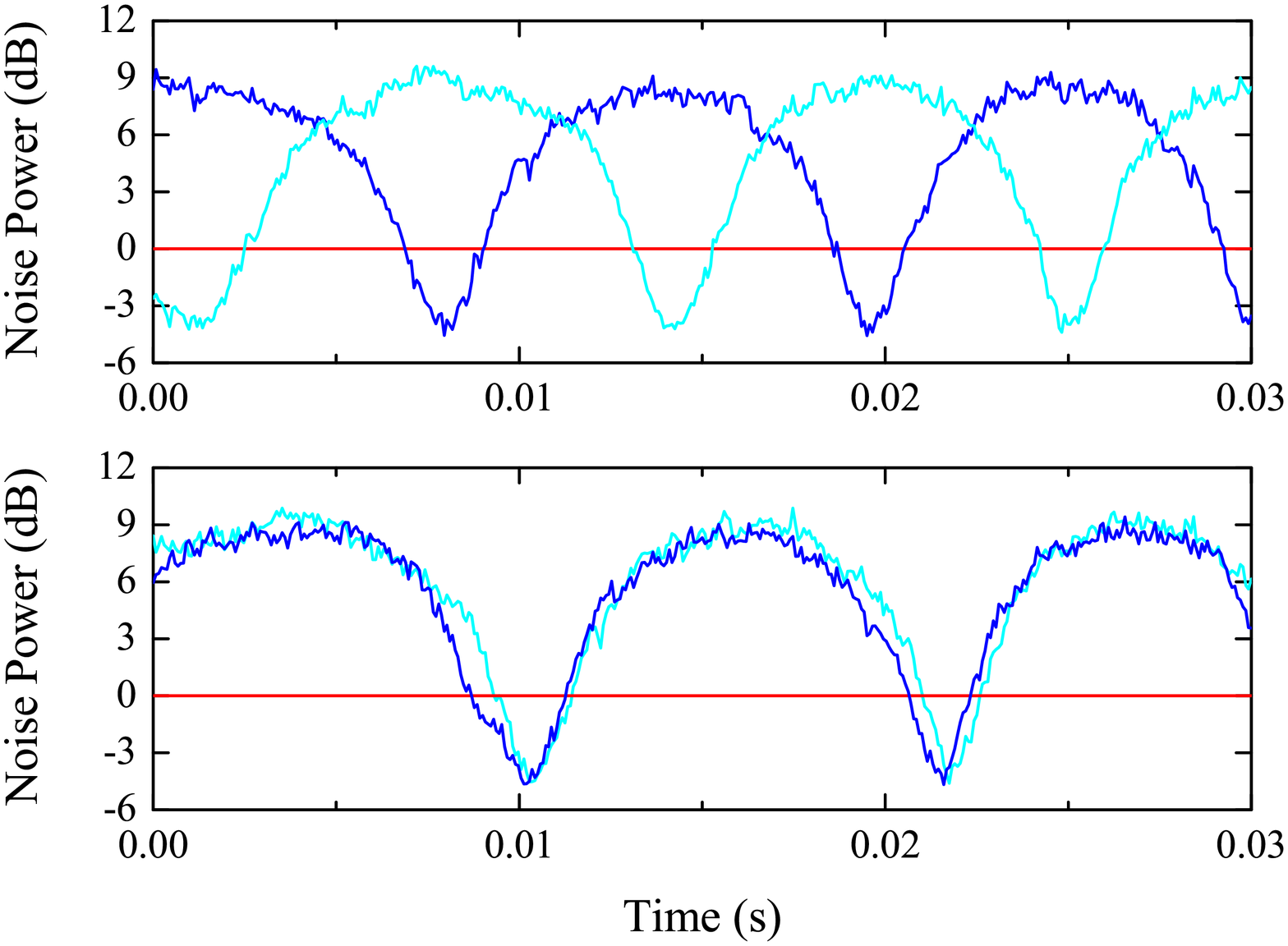}%
\caption{\label{verysmallscan}Normalized noise variances at 3.5
MHz of the $\pm 45^{\circ}$ modes while scanning the local
oscillator phase. The first plot corresponds to in-phase homodyne
detections and the second one in-quadrature. Squeezing is well
observed on orthogonal quadratures. The resolution bandwidth is
set to 100 kHz and the video bandwidth to 1 kHz.}
\end{figure}

In a first series of experiments, the plate angle is adjusted to
be almost zero. This tuning can be done by looking at the
individual noises which should be in that case phase-insensitive.
Squeezing of the rotated modes is thus expected on orthogonal
quadratures, as it is well-known for a standard OPO. Typical
spectrum analyser traces while scanning the local oscillator phase
are shown on Fig. \ref{verysmallscan}. Normalized noise variances
of the $\pm 45^{\circ}$ vacuum modes at a given noise frequency of
3.5 MHz are superimposed for in-phase and in-quadrature homodyne
detections. One indeed observes, as expected, correlations and
anti-correlations of the emitted modes on orthogonal quadratures.

\begin{figure}[htpb!]
\includegraphics[width=.95\columnwidth,clip=]{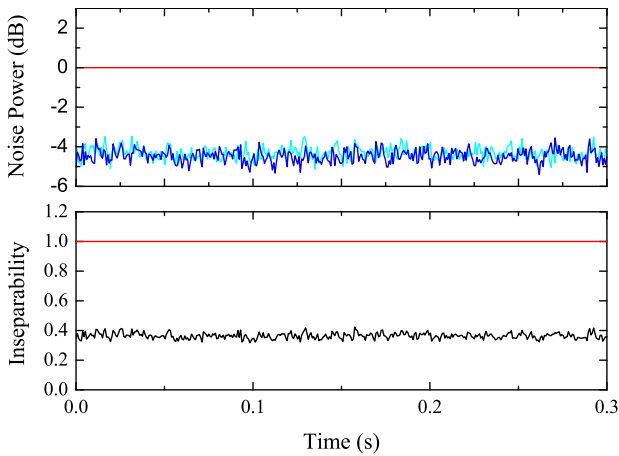}%
\caption{\label{lock}Normalized noise variances at 3.5 MHz of the
$\pm 45^{\circ}$ modes and inseparability criterion for signal and
idler modes. The homodyne detections are in-quadrature. After
correction of the electronic noise, the inseparability criterion
reaches $0.33\pm 0.02$. The resolution bandwidth is set to 100 kHz
and the video bandwidth to 300 Hz.}
\end{figure}

Figure \ref{lock} gives the simultaneous measurement of the noise
reductions for a locked local oscillator phase. $-4.3 \pm 0.3$ dB
and $-4.5 \pm 0.3$ dB below the standard quantum limit are
obtained for the two rotated modes. After correction of the
electronic noise, the amounts of squeezing reach $-4.7 \pm 0.3$ dB
and $-4.9 \pm 0.3$ dB. These values have to be compared to the
theoretical value expressed in Eq. (\ref{without}). By taking
$\sigma=0.9$, $\Omega=0.1$, $\kappa=0.025$ and $\kappa'=0.03$, the
expected value before detection is $-7.5$ dB. The detector quantum
efficiency is estimated to $0.95$, the fringe visibility is $0.97$
and the propagation efficiency is evaluated around $0.99$. These
values give an overall detection efficiency of $0.95 \cdot
0.97^{2} \cdot 0.99=0.88$. After detection, the expected squeezing
is thus reduced to $-5.5$ dB. The small discrepancy with the
experimental values can be due to the presence of walk-off which
limits the modes overlap, a critical point for two-mode squeezing.

From the electronic noise corrected squeezing values, one can
infer the Duan and Simon inseparability criterion defined as the
half sum $\Delta$ of the squeezed variances \cite{duan,simon}. For
a symmetric gaussian two-mode state, this criterion is a necessary
and sufficient condition of non-separability. We obtained a value
of $\Delta=0.33 \pm 0.02$ well below the unit limit for
inseparability. It is worth noting that the simultaneous double
homodyne detection permits a direct and instantaneous verification
of this criterion by adding the two squeezed variances.

The EPR criterion is related to an apparent violation of a
Heisenberg inequality \cite{reid}: the information extracted from
the measurement of the two quadratures of one mode provides values
for the quadratures of the other mode that violate the Heisenberg
inequality. This criterion is related to the product of
conditional variances: $V_{P_1|P_2}\,V_{Q_1|Q_2}<1$ where $P_i$ et
$Q_i$ are two conjugate quadratures and $V_{X_1|X_2}$ the
conditional variance of $X_1$ knowing $X_2$. The knowledge of the
previous squeezed quadratures and of the individual noise of the
entangled modes give the conditional variances. The noise of
signal and idler modes are phase-insensitive and reach $8.2\pm
0.5$ dB above shot noise (fig. \ref{individualscan}). We obtained
thus a product of conditional variances equal to $0.42\pm 0.05$,
which confirms the EPR character of the measured correlations.

The entanglement can be quantified by the entropy of formation --
or entanglement of formation $EOF$ -- for symmetric gaussian
states introduced in Ref.\cite{giedke}, which represents the
amount of pure state entanglement needed to prepare the entangled
state. This entropy can be directly derived from the
inseparability criterion value $\Delta$ by
\begin{eqnarray}
EOF=c_{+}\log_{2}(c_{+})-c_{-}\log_{2}(c_{-})
\end{eqnarray}
with
\begin{eqnarray}
c_{\pm}=(\Delta^{-1/2} \pm \Delta^{1/2})^{2}/4
\end{eqnarray}
From this expression, we calculate an entanglement of formation
value of $EOF=1.1 \pm 0.1\,ebits$. To the best of our knowledge,
our setup generates the best EPR entangled beams to date produced
in the continuous variable regime. Let us note that such a degree
of entanglement should correspond to a fidelity equal to $0.75$ in
a unity gain teleportation experiment.

This non-classical behavior exists also without the plate and we
have obtained in that case almost the same degree of entanglement.
The first experimental demonstration of continuous variable EPR
entanglement was obtained with such a type II OPO below threshold
\cite{Ou}. However, the linear coupling -- even for a plate
rotated by a very small angle -- can make easier the finding of
experimental parameters for which entanglement is observed.
Furthermore, the degenerate operation with bright beams above
threshold makes possible to match the homodyne detection without
infrared injection of the OPO.

The entanglement of the generated two-mode state is preserved for
very low noise frequencies, down to 50 kHz. In the experimental
quantum optics field, non-classical properties are generally
reported in the MHz range -- as it is the case in this paper up to
now -- due to large classical excess noise at lower frequencies.
Experimental details and possible applications of this low
frequency results are reported in \cite{laurat04b}.

\begin{figure}[htpb!]
\includegraphics[width=.95\columnwidth,clip=]{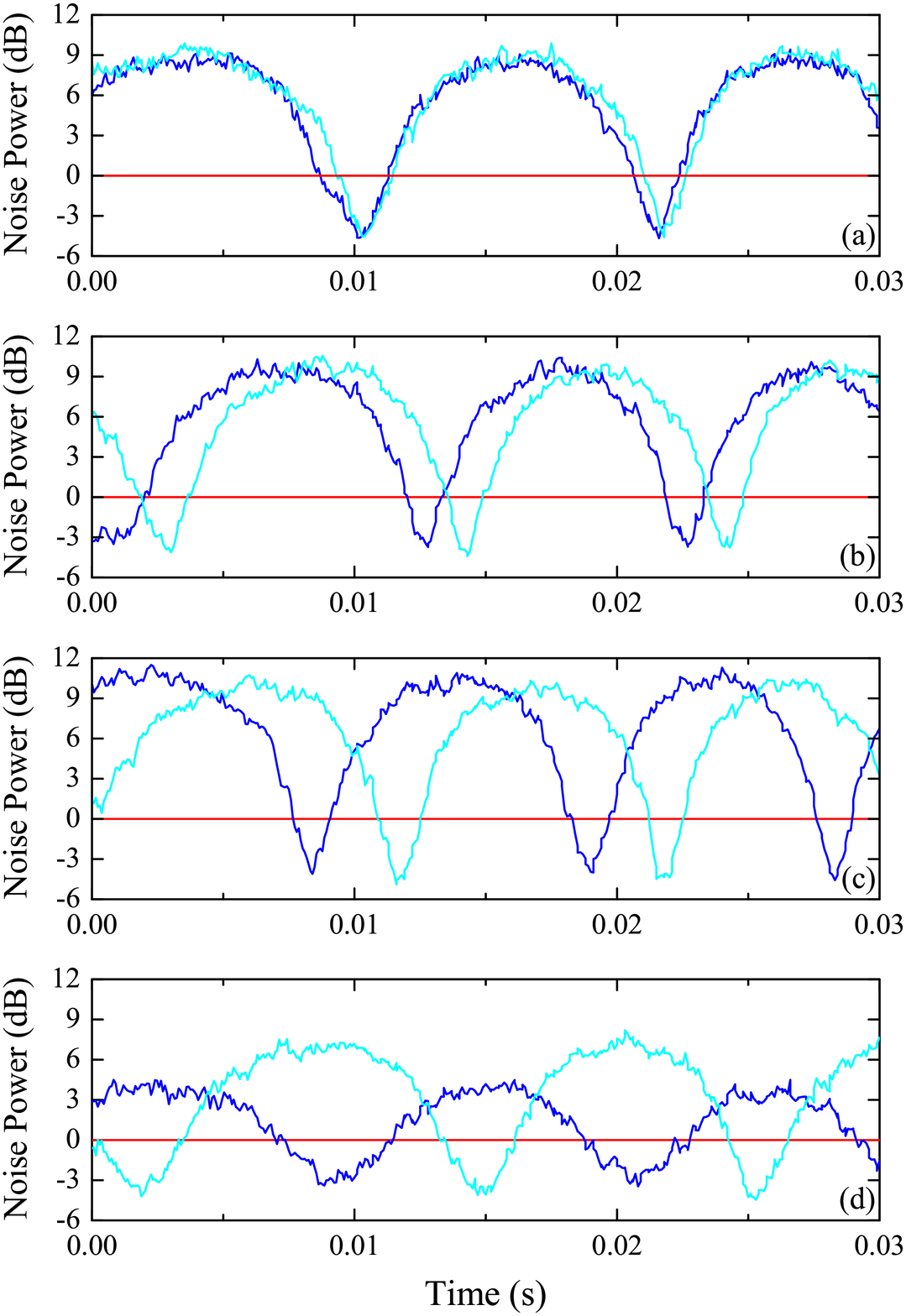}%
\caption{\label{scan}Normalized noise variances at 3.5 MHz of the
$\pm 45^{\circ}$ modes while scanning the local oscillator phase
for different coupling parameters: (a) $c=0$, (b) $c=0.35$, (c)
$c=0.85$ and (d) $c=1.8$. Dark lines correspond to the $A_{-}$
mode and light ones to the $A_{+}$ mode. The homodyne detections
are in-quadrature. The resolution bandwidth is set to 100 kHz and
the video bandwidth to 1 kHz.}
\end{figure}

\begin{figure*}[htpb!]
\includegraphics[width=1.8\columnwidth,clip=]{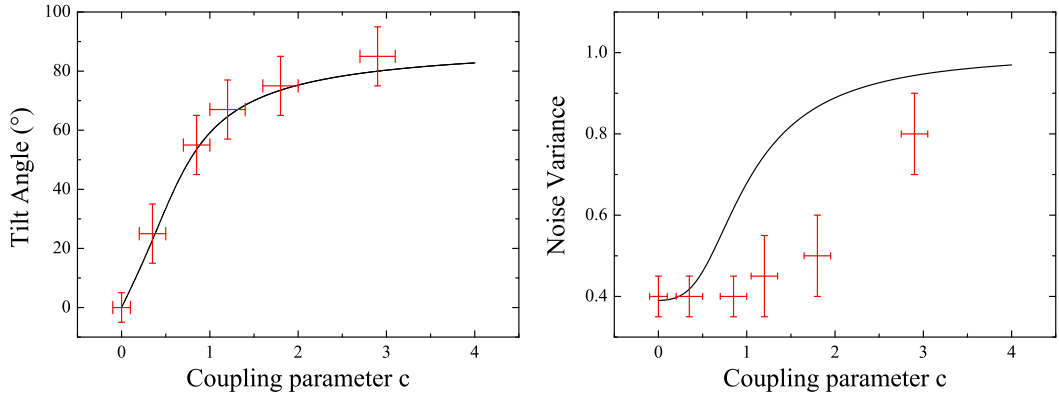}%
\caption{\label{resume} Measured tilt angle $\theta$ and noise
variance as a function of the coupling parameter c, for the $A_-$
mode. The black lines give theoretical predictions calculated from
Eq. (\ref{theta}) and (\ref{amoins1}) for $\sigma=0.9$ and
$\Omega=0.1$.}
\end{figure*}

\subsection{Results as a function of the linear coupling}
The two-mode state generated by the self-phase-locked OPO is then
characterized for different angles of the plate. We use the
in-quadrature setup of the homodyne detections for which squeezing
on orthogonal quadratures is observed simultaneously on the
triggered spectrum analyzers. When the coupling increases, the
squeezing is not obtained on orthogonal quadratures anymore.
Figure \ref{scan} gives for four increasing coupling parameters
the normalized noise variances of the rotated modes while scanning
the local oscillator phase. In Fig. \ref{resume}, we give the
experimentally measured tilt angle $\theta$ and associated noise
variance as a function of the coupling parameter $c$. One can
check on the figure the validity of the theoretical expression of
$\theta$ given in Eq. (\ref{theta}). We observe that the squeezing
of the $A_{-}$ mode decreases but more slowly than expected. We
also note that the squeezing of the $A_{+}$ mode slightly
decreases while this noise reduction is theoretically independent
of the coupling.

\begin{figure}[htpb!]
\includegraphics[width=.94\columnwidth,clip=]{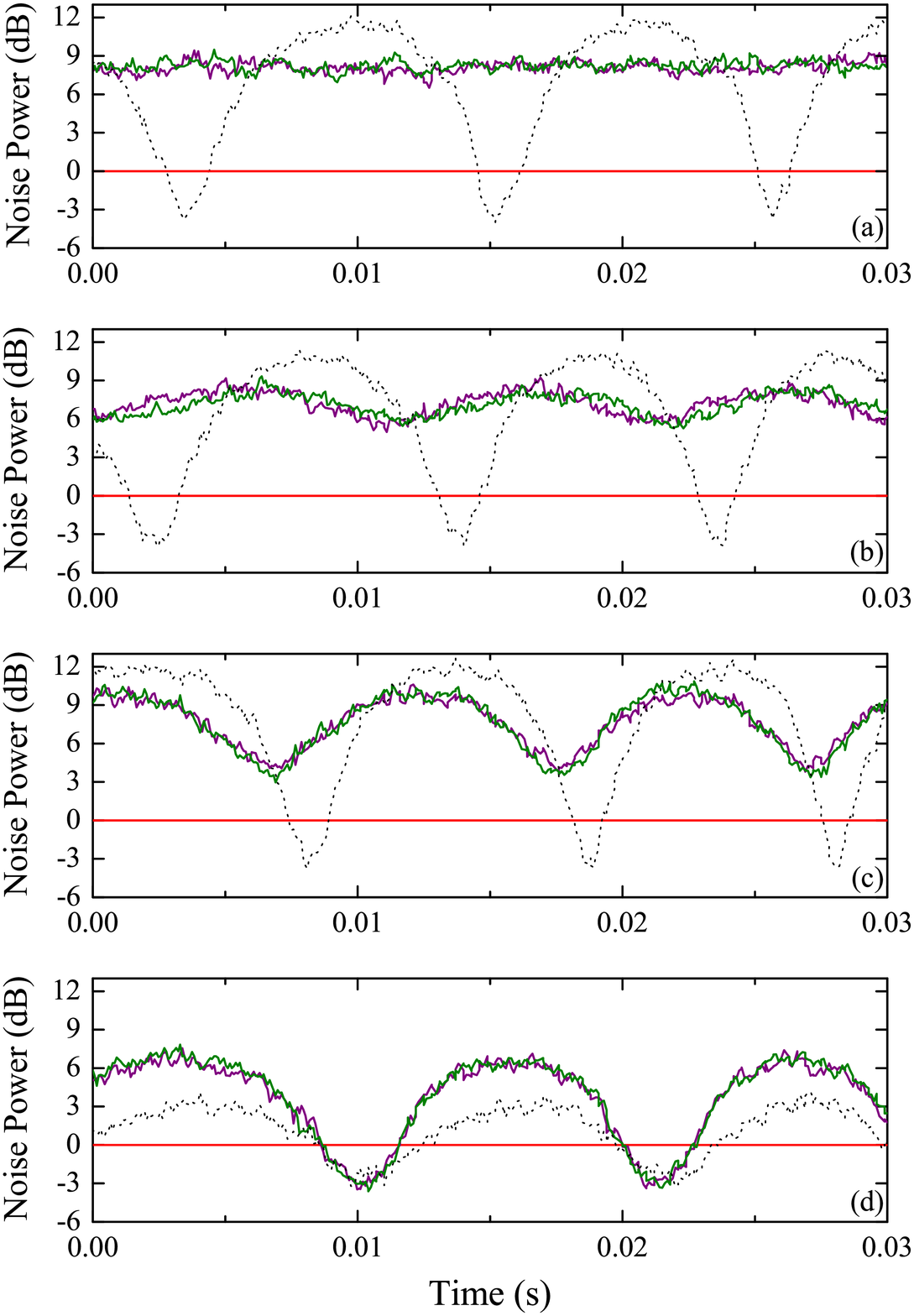}%
\caption{\label{individualscan}Normalized noise variances at 3.5
MHz of the signal and idler modes while scanning the local
oscillator phase for different coupling parameters: (a) $c=0$, (b)
$c=0.35$, (c) $c=0.85$ and (d) $c=1.8$. The black dotted lines
correspond to the noise power on the difference of the two modes.
The homodyne detections are in-phase. The resolution bandwidth is
set to 100 kHz and the video bandwidth to 1 kHz.}
\end{figure}

The noise of the signal and idler modes also depends on the
presence of the plate as demonstrated in Sec. \ref{sec:theory}.
The individual noises become phase-sensitive, and even squeezed
below the standard quantum limit, when the coupling increases.
Figure \ref{individualscan} gives the phase dependance of the
signal and idler modes for the same four coupling parameters than
in Fig. \ref{scan}.

\subsection{Optimization of EPR entanglement by polarization adjustement}
When the plate is rotated, squeezing is not observed on orthogonal
quadratures anymore. Thus, as shown in Sec.\ref{sec:theory}, the
EPR entanglement is not the maximal available one. In order to
extract the maximal entanglement, one has to perform a phase-shift
of the $A_{+}$ and $A_{-}$ modes. Such an arbitrary phase-shift
can be done thanks to a couple of a $\lambda/2$ and a $\lambda/4$
plates added at the output of the OPO (Fig.\ref{setup}).

\begin{figure}[htpb!]
\includegraphics[width=.95\columnwidth,clip=]{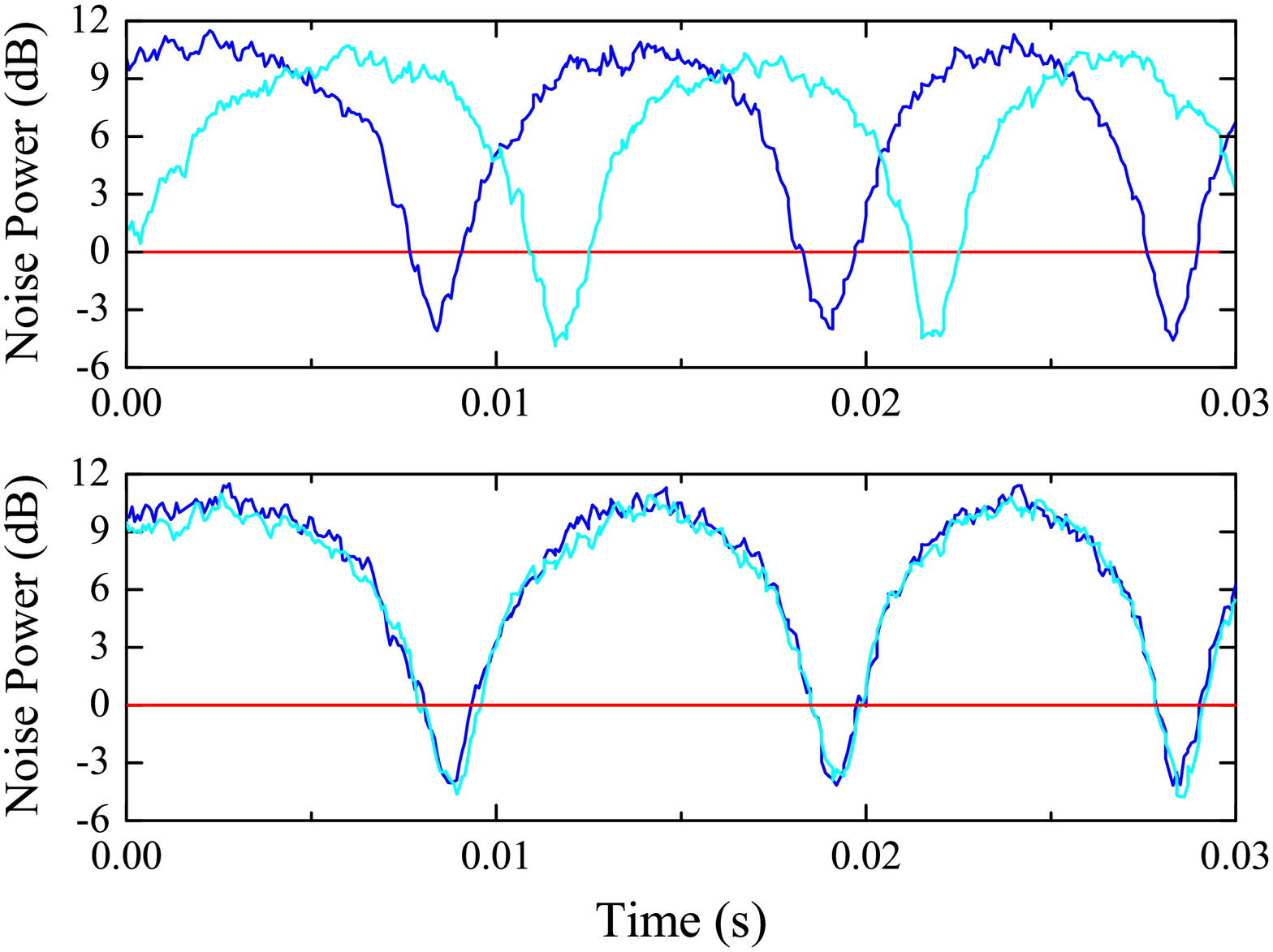}%
\caption{\label{corrige}Normalized noise variances at 3.5 MHz of
the rotated modes while scanning the local oscillator phase for a
coupling parameter $c=0.35$, before and after the non-local
operation. The homodyne detections are in-quadrature. After this
operation, squeezing is observed on orthogonal quadratures. The
resolution bandwidth is set to 100 kHz and the video bandwidth to
1 kHz.}
\end{figure}

Figure \ref{corrige} gives the normalized noise variances of the
rotated modes for a coupling parameter $c=0.35$, before and after
the phase-shift. The homodyne detections are operated in
quadrature so that squeezing on orthogonal quadratures is observed
simultaneously on the spectrum analyzers. After the operation
performed, squeezing is obtained on orthogonal quadratures as in a
standard  type II OPO without coupling.

\section{Conclusion}
\label{sec:conclusion}A self-phase-locked type II OPO associates
to the usual non-linear coupling between the signal and idler
modes a linear mixing by the way of a rotated quarter-wave plate
inserted inside the optical cavity. We have demonstrated
theoretically and confirmed experimentally that this original
device generates a two-mode non-classical state that exhibits a
very rich and interesting behavior in terms of squeezing and
correlation properties. Quantum correlations and anti-correlations
of the signal and idler modes are obtained on non-orthogonal
quadratures depending on the angle of the plate. Furthermore, by a
suitable change of polarization, the entanglement can be maximized
and put into standard form, i.e. correlations and
anti-correlations on orthogonal quadratures. The observed
entanglement has been characterized in terms of covariance matrix
and logarithmic negativity.

The experimental investigation of this original device required
the setup of two simultaneous homodyne detections. We have
detailed the operation of the system as a function of the coupling
parameter -- for the signal and idler modes which are entangled as
well for the $\pm 45^{\circ}$ rotated modes which have squeezed
fluctuations -- and found the experimental behavior consistent
with the theory. In the case of a very small coupling, we have
reported what is to our knowledge the best entangled beams ever
produced in the continuous variable regime. A value of the
inseparability criterion as low as $0.33 \pm 0.02$, well below the
limit of unity, is obtained. This entanglement corresponds to a
value of the entanglement of formation of $1.1 \pm 0.1$ ebits. We
also achieved EPR entanglement and squeezing at very low noise
sideband frequencies down to 50 kHz \cite{laurat04b}.

The next step is the characterization of the quantum properties of
this system operated above threshold. The linear coupling induced
by the plate results in a phase-locking of the signal and idler
fields at frequency degeneracy, which permits to access the phase
fluctuations of the bright twin beams. The predicted quantum
properties of the system are similar and should open the
possibility to generate bright entangled beams \cite{longcham2}.
Up to now, phase-locking and intensity correlations below the
standard quantum limit have been observed but the phase
anticorrelations are still slightly above shot noise
\cite{laurat04d}. Improvements of the setup are currently in
progress.

\begin{acknowledgments}
Laboratoire Kastler-Brossel, of the Ecole Normale Sup\'{e}rieure
and the Universit\'{e} Pierre et Marie Curie, is associated with
the Centre National de la Recherche Scientifique (UMR 8552).
Laboratoire Mat{\'e}riaux et Ph{\'e}nom{\`e}nes Quantiques is a F{\'e}d{\'e}ration de
Recherche (CNRS FR 2437). This work has been supported by the
European Commission project QUICOV (IST-1999-13071) and ACI
Photonique (Minist\`ere de la Recherche et de la Technologie).
\end{acknowledgments}

\end{document}